\documentclass[doublecol]{epl2}
\usepackage{graphicx}
\usepackage[T1]{fontenc}
\title{Charge and spin density in the helical Luttinger liquid}
\shorttitle{Charge and spin density in the helical Luttinger liquid} %Insert here a short version of the title if it exceeds 70 characters

\author{N. Traverso Ziani\inst{1} \and C. Fleckenstein\inst{1} \and F. Cr\'epin\inst{2} \and B. Trauzettel\inst{1,3}}
\shortauthor{N. Traverso Ziani \etal}

\institute{
\inst{1}Institute for Theoretical Physics and Astrophysics, University of W\"{u}rzburg, 97074 W\"{u}rzburg, Germany  \\
\inst{2}Laboratoire de Physique Th\'eorique de la Matière Condens\'ee, UPMC, CNRS UMR 7600, Sorbonne Universit\'es, 4 place Jussieu, 75252 Paris Cedex 05, France\\
\inst{3}{Department of Physics, University of California, Berkeley, California 94720, USA}}
\pacs{71.10.Pm}{Fermions in reduced dimensions (anyons, composite fermions, Luttinger liquid, etc.)}
\pacs{73.20.Qt}{Electron solids}

\abstract{The weakly interacting helical Luttinger liquid, due to spin momentum locking, is characterized by extremely peculiar local observables: we show that the density-density correlation functions do not exhibit signatures of Friedel and Wigner oscillations, and that spin-spin correlation functions, which are strongly anisotropic, witness the formation of a planar spin wave. Moreover, we demonstrate that the most relevant scattering potentials involving a localized impurity are not able to modify the electron density, while only magnetic impurities can pin the planar spin density wave.}
\begin{document}

\maketitle
\section{Introduction}
\noindent One dimensional gapless quantum systems are characterized by a peculiar behavior: if (arbitrarily weak) inter-particle interaction is present, the low energy excitations have a collective nature and are described by the Luttinger liquid theory\cite{haldane,voit,giamarchi}. Experimental evidence of the validity of the Luttinger liquid theory in fermionic systems has been achieved in a number of different devices, ranging from carbon nanotubes\cite{cnt1,cnt2} and quantum wires\cite{wire1,wire2}, to Bechgaard salts\cite{salts} and edges of quantum Hall bars\cite{hall1,hall2,hall3}. Exotic behaviors related to the Luttinger liquid, like spin charge separation\cite{spincharge} and charge fractionalization\cite{fractionalization1,fractionalization2}, have also been observed. Moreover the Luttinger liquid picture is able to capture the low energy properties of gapless one dimensional bosonic\cite{bosons} and spin systems\cite{spin}. In addition to its wide range of applicability, the Luttinger liquid picture has another important technical advantage: its Hamiltonian describes a collection of free bosons, and the fundamental field operators (right and left movers) of the theory can also be expressed by means of the creation and annihilation operators of such bosons, so that the correlation functions can very often be calculated analytically\cite{voit}. As a direct consequence of its wide range of validity, the relation between the fundamental field operators of the Luttinger liquid and the observables do depend on the system under consideration. Even though the thermodynamics of all the Luttinger liquid pictures of various systems is the same due to the formal analogy of the Hamiltonians, the physical behaviors can hence be completely different. A striking example of Luttinger liquid with exotic properties is the helical Luttinger liquid\cite{helical1,helicalplus} (hLL). It is characterized by spin momentum locking: particles with spin up and down with respect to a well defined $z$ axis propagate in opposite directions. The hLL has been studied in details since it describes the edges of two dimensional topological insulators (2DTI)\cite{hasan,qi,moore,bernevig,km1,km2,koenig7,altro} and, to some extent, spin-orbit coupled quantum wires in the presence of a magnetic field\cite{soc1,soc2,soc3,soc4}. The interest in such systems is motivated by potential applications in spintronics\cite{st1,st2,st3} and quantum computation\cite{qc1,qc2}. The properties under consideration are most often related to transport, partially because the explanation of the weak temperature dependence of the non-perfectly quantized conductance in long samples still represents an open and challenging issue, even though several scattering mechanisms have been inspected\cite{tr1,tr2,tr3,tr4,tr6,tr7,tr8,tr9,tr10,tr11,tr12}. Much less emphasis has on the other hand been devoted to its local properties. Recently, it has been demonstrated that in the strong interaction regime a very peculiar state, characterized by fractional oscillations, emerges\cite{prl,cav}.\\
In this Letter we concentrate on the weak interaction regime: first we investigate the local density-density and spin-spin correlation functions and we find that they are very different from the ones characterizing usual one-dimensional quantum wires (QW): the density-density correlation function of the hLL is a non oscillating function, while in QW one finds clear signatures of Friedel oscillations; the spin-spin correlation functions are strongly anisotropic and witness the formation of a spin density wave on the plane perpendicular to the $z$ axis, while in QW the spin density waves are isotropic. Afterwards we address the problem of how a localized impurity influences the electron and spin densities of the hLL. We choose the three impurity potentials that have been demonstrated to be the most relevant in the renormalization group sense: one-particle backscattering, allowed only if time reversal symmetry is broken, two particle backscattering, and a certain type of inelastic backscattering. We demonstrate that, to all orders in the perturbative expansion in the impurity potentials, the electron density remains unchanged, while one particle backscattering can pin the planar spin density wave.\\
The outline of the Letter is the following: we start by describing the model we adopt for the hLL, and we study the density-density and the spin-spin correlation functions at zero temperature. Afterwards we address the effect on the charge and spin densities of a localized impurity.
\section{Model and one-particle observables}
\label{sec:model}
The model consists of the usual hLL harmonic Hamiltonian $H_0$ describing the helical edge.\\
After imposing periodic boundary conditions on a segment of length $L$, one has $H_0=\int_0^L h_0 dx$, with
\begin{equation}
h_0=\frac{\prod}{2}+\frac{\left(\partial_x\phi\right)^2}{2},\label{eq:h_0}
\end{equation}
where $\phi$ is the usual Luttinger bosonic field and $\prod$ is canonically conjugated to $\phi$. The Luttinger fermionic fields $\psi_{\pm}$ for right and left moving electrons are related to the bosonic fields by the bosonization identity
\begin{equation}
\psi_{\pm}(x)=\frac{\mathcal{F}_\pm}{\sqrt{2\pi\alpha}}e^{\pm ik_Fx}e^{i\frac{\theta(x)}{\sqrt{\pi K_L}}\pm i\sqrt{\pi K_L} \phi(x)},
\end{equation}
where $\mathcal{F}_\pm$ are the Klein factors for particles with spin $\pm$\cite{voit}, $K_L$ is the Luttinger parameter, with $K_L<1$ for repulsive interaction, and $K_L=1$ for free electrons, $k_F$ is the Fermi momentum, $\alpha\sim 1/k_F$ is a cutoff length, and $\theta(x)$ is given by $\partial_x\theta/\pi=\prod$. Moreover we set $\hbar=v_F=1$, with $v_F$ the Fermi velocity. In this Letter we assume weak interaction $1/2<K_L<1$ since we are not interested in correlated two particle backscattering\cite{helical1}.
The relation between the fields $\psi_{\pm}$ and the physical electrons is unusual: due to spin momentum locking the Fermi spinor $\Psi(x)$ for physical particles becomes\cite{helical1}
\begin{equation}
\Psi(x)=(\psi_+(x),\psi_-(x))^T,\label{eq:fermioperator2}
\end{equation}
so that, for some appropriate $z$ axis spin up(down) electrons propagate in the right(left) direction. The electron density operator in terms of the bosonic operators reads
\begin{eqnarray}
\rho(x)&=&\Psi^\dag(x)\Psi(x)=\frac{2k_F}{\pi}+\frac{\sqrt{K_L}\partial_x\phi(x)}{\sqrt{\pi}}\\
&=&\frac{2k_F}{\pi}+\frac{\Delta N_++\Delta N_-}{L}+\rho_{\mathrm{bos}}.
\end{eqnarray}
Evidently only the long wave component is present and the Friedel term, that characterizes usual Luttinger liquids, is absent. In the second line we have separated the contribution to the density arising from the excess number of electrons with spin up(down), $\Delta N_+$($\Delta N_-$), over $k_F$ and the fluctuations $\rho_{\mathrm{bos}}$, which are linear in the bosonic operators $b_n$ ($n\in Z$, $n\neq 0$), and read as
\begin{equation}
{\rho_{\mathrm{bos}}=\frac{\sqrt{K_L}}{L}\sum_{n\neq 0}|n|^{\frac{1}{2}}e^{-\frac{\alpha\pi |n|}{L}}e^{-\frac{2in\pi x}{L}}\left(b^\dag_n+b_{-n}\right).}
\end{equation}
The spin density, defined as
\begin{equation}
\mathbf{s}(x)=\Psi^\dag(x){\mathbf{\sigma}}\Psi(x),
\end{equation}
where $\mathbf{\sigma}=(\sigma^1,\sigma^2,\sigma^3)$ is the vector of Pauli matrices, is also very different from the one describing QW systems. Explicitly it is given by
\begin{eqnarray}
s_1(x)&=&(\psi^\dag_+(x)\psi_-(x)+\mathrm{h.c.}),\\
s_2(x)&=&-i(\psi^\dag_+(x)\psi_-(x)-\mathrm{h.c.}),\\
s_3(x)&=&(\psi^\dag_+(x)\psi_+(x)-\psi^\dag_-(x)\psi_-(x)).
\end{eqnarray}
Since we adopt periodic boundary conditions the amount of information that can be gained by the inspection of the average charge spin density is very limited, since they are both constant. We hence concentrate on correlation functions, that can be evaluated exactly by means of the bosonization technique.
\section{Correlation functions}
The density-density correlation function $c_1(x)$ is defined as
\begin{equation}
c_1(x)=_N\langle 0|\rho(x)\rho(0)|0\rangle_N,
\end{equation}
where $|0\rangle_N$ is the ground state of $H_0$ for ${N=\frac{k_F L}{\pi}+N_++N_-}$ particles in the system. Its explicit expression can be written as
\begin{equation}
c_1(x)=\frac{N^2}{L^2}+\frac{K_L}{L^2}\left[\frac{1-z(x)}{z(x)^2}+\mathrm{h.c.}\right],
\end{equation}
with
\begin{equation}
z(x)=1-e^{-2\pi\frac{\alpha}{L}}e^{2i\pi \frac{x}{L}}.\label{eq:dd}
\end{equation}
These results show that the $2k_F$ Friedel oscillations of the density-density correlation function are absent in the hLL. This behavior is due to the topological nature of the system, that forbids one particle backscattering if time reversal symmetry is not broken. Moreover note that, even for small $N\sim3$ the second addend in $c_1(x)$ is negligible with respect to the first for any $\alpha<x<L-\alpha$. For other values of $x$ the theory is in any case unreliable since $\alpha$ is the cutoff length. We will hence, from now on, adopt the approximation $c_1(x)\sim N^2/L^2$.\\

The spin-spin correlation functions $s_{ij}(x)$, are defined as
\begin{equation}
s_{ij}(x)=\frac{2L^2}{N^2}{_N\langle 0|s_i(x)s_j(0)|0\rangle_N}.
\end{equation}
Considering again only $\alpha<x<L-\alpha$, one finds $s_{13}=s_{31}=s_{23}=s_{32}=0$;
\begin{eqnarray}
s_{11}\!&=&\!s_{22}\!=\!-\left|\frac{z(0)}{z(x)}\right|^{2K_L}{\cos\left[\frac{2\pi N x}{L}-\rm{arg}\left[\frac{z(x)}{z(-x)}\right]\right]\nonumber}\\
s_{12}\!&=&\!-s_{21}\!=\!-\left|\frac{z(0)}{z(x)}\right|^{2K_L}{\sin\left[\frac{2\pi N x}{L}-\rm{arg}\left[\frac{z(x)}{z(-x)}\right]\right],\nonumber}
\end{eqnarray}
and finally, along the axis defining the spin momentum locking
\begin{equation}
s_{33}(x)=\frac{1}{K_L^2}\left[c_1(x)-\frac{N^2}{L^2}\right]\sim 0\,\,\,\,\,\, \mathrm{for}\,\,\,x>\alpha.
\end{equation}
The spin component in the direction of the $z$ axis is hence featureless, while the correlations in the $xy$ plane are nontrivial. Moreover, as shown in Fig.1, the oscillations become more pronounced and decay slower as $K_L$ becomes smaller. Note that the oscillations are more pronounced than in the usual QW, where an interaction independent damping is present due to the correlations in the spin field\cite{giamarchi}.\\
\begin{figure}[htbp]
\begin{center}
\includegraphics[width=8.5cm,keepaspectratio]{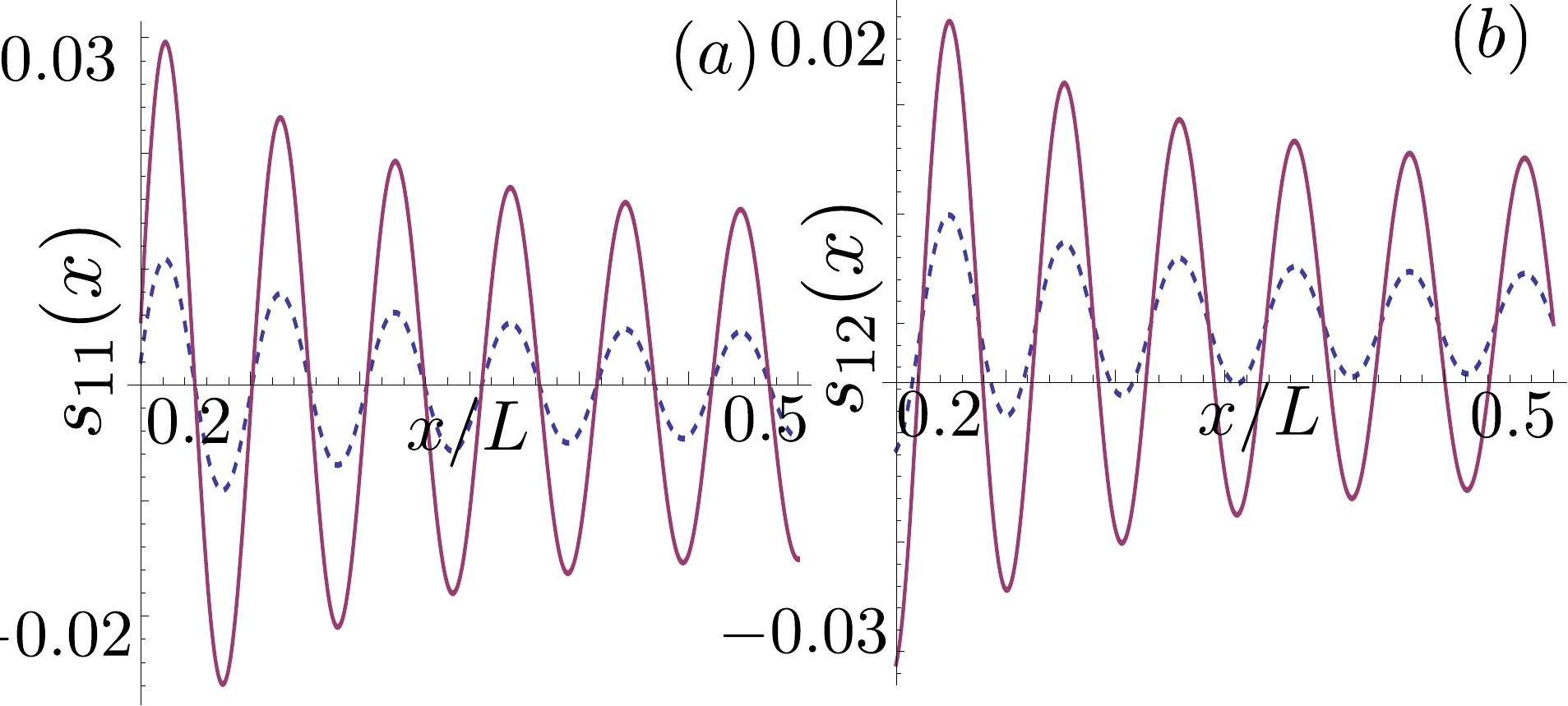}
\caption{(Color online) The spin-spin correlation functions $s_{11}(x)$ (panel (a)) and $s_{12}(x)$ (panel (b)) as a function of $x/L$ for $N=20$ and $K_L=0.9$, blue (dotted) lines and $K_L=0.7$ purple (plain) line.}
\label{fig:fig1}
\end{center}
\end{figure}
To better understand the meaning of the oscillations in $s_{ij}(x)$ ($i,j=1,2$) it is useful to consider the spin resolved density-density correlation functions. In particular we address the probability density $p_{\gamma,\delta}(x)$, normalized to $N^2$, of finding an electron at position $x$ with spin projection $\gamma=\pm$ along the $\delta=1,2=x,y$ axis if an electron with spin + along the $x$ axis is present. One simply has
\begin{equation}
p_{\gamma,\delta}(x)=\left(c_1(x)+\frac{N^2}{2L^2}\gamma s_{1\delta}(x)\right).
\end{equation}
\begin{figure}[htbp]
\begin{center}
\includegraphics[width=8.5cm,keepaspectratio]{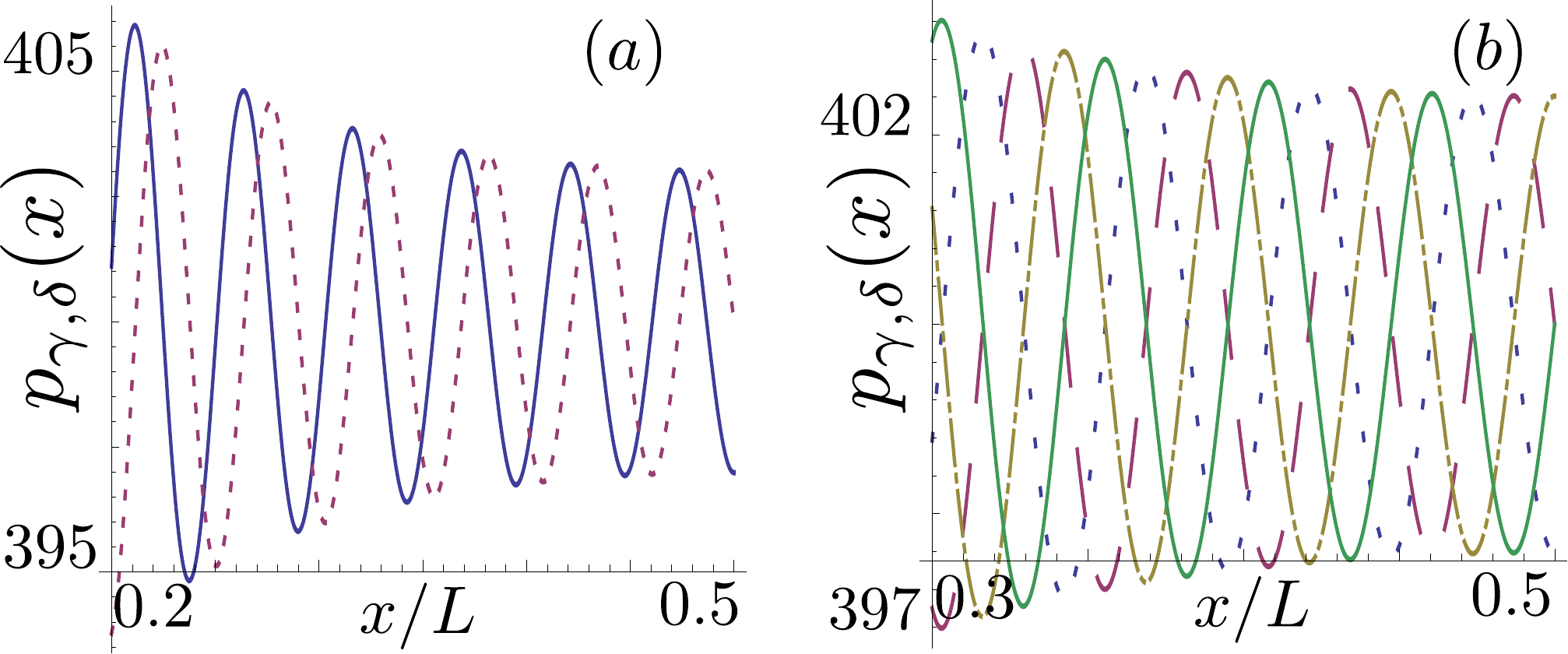}
\caption{(Color online)(a) $p_{+,1}(x)$ in blue (plain) and $p_{+,2}(x)$ in purple (dotted) as a function of $x/L$; (b) $p_{+,1}(x)$ in blue (dotted), $p_{+,2}(x)$ in purple (dashed), $p_{-,1}(x)$ in brown (dash-dotted) and $p_{-,2}(x)$ in green (plain). In both panels $K_L=0.7$, $N=20$.}
\label{fig:fig2}
\end{center}
\end{figure}
Note that a slight deviation (of order $1/N^2$) from the correct normalization is expected due to the impossibility of describing the regime $x<\alpha$ within the Luttinger liquid theory. The results are shown in Fig.2 for $N=20$ and $K_L=0.7$. The spin density rotates in the $xy$ plane while moving along the edge in such a way that it forms a planar helix. This behavior is in sharp contrast to the case of usual quantum wires where the non-diagonal spin-spin correlation functions vanish. Note that a signature of this peculiar spin order had already been predicted in magnetically defined quantum dots\cite{giacomo1,giacomo2}, where, however, translational invariance is broken and the helix is already visible at the level of one particle operators. Moreover in that case magnetic impurities were added to the system, while here we demonstrate that such a spin helix is a property of the unperturbed hLL, and the role of magnetic impurities is to pin them.
\section{Effect of a localized impurity}
\label{sec:impurity}
We now consider the hLL, with Hamiltonian $H_0$, in the presence of three possible impurity-induced backscattering processes: one-particle backscattering $H_{1}$, allowed only if time reversal invariance is broken, as in the case of magnetic impurities, two-particle backscattering $H_{2}$, which does not break time reversal symmetry but requires electron-electron interaction, and a certain type of inelastic backscattering $H_{3}$, which also requires electron-electron interaction and consists of taking one electron from one branch to the other and creating a particle-hole excitation. This term has been argued to be important in generic helical liquids\cite{helical2,generalhelical}. The reason for the choice of these three processes is simplicity. They have been shown to be the lowest dimensional allowed scattering terms\cite{helical2} in a renormalization group sense. Explicitly,
\begin{eqnarray}
H_{1}&=&v_1\left(\psi_+^\dag\psi_-\right)_{x=0}+\mathrm{h.c.}\label{Eq:H_1}\\
H_{2}&=&v_{2}\left(\psi_+^\dag\partial_x\psi_+^\dag\psi_-\partial_x\psi_-\right)_{x=0}+\mathrm{h.c.}\label{Eq:H_{2}}\\
H_{3}&=&v_{3}\left[\!\left(\!\partial_x\psi_-^\dag\psi_-\!-\!\psi_+^\dag\partial_x\psi_+\!\right)\psi^\dag_-\psi_+\!\right]_{x=0}\!+\!\mathrm{h.c.}\label{Eq:H_{3}}
\end{eqnarray}
where $v_i$ ($i=1,2,3$) are free parameters. In usual Luttinger liquids the effect of $H_i$ ($i=1,2,3$)  is already visible at the level of the electron density: the impurity breaks translational invariance and pins oscillations. For example $H_1$ induces Friedel oscillations, with wavevector $2k_F$. In hLL the situation is different due to the absence of Friedel oscillations.\\
In order to evaluate the influence of $H_i$ ($i=1,2,3$) on the electron density we compute the average electron density $\langle\rho(x)\rangle_i$ at temperature $T$, given by
\begin{equation}
\langle\rho(x)\rangle_i=\frac{\mathrm{Tr}\left\{e^{-\beta({H_0+H_i})}\rho(x)\right\}}{\mathrm{Tr}\left\{e^{-\beta({H_0+H_i})}\right\}},\label{eq:average}
\end{equation}
where $\mathrm{Tr}\{\cdot\}$ indicates the trace over a basis of the Hilbert space and $\beta=1/(k_BT)$, where $k_B$ is the Boltzman constant and $T$ the temperature. The calculation can be performed to all orders in perturbation theory. In order to do so it is convenient to change into the (imaginary time) interaction picture, that is to define, for any operator $O$, its interaction picture counterpart $O_{\mathrm{INT}}(\tau)$ as
\begin{equation}
O_{\mathrm{INT}}(\tau)=e^{H_0\tau}Oe^{-H_0\tau}.
\end{equation}
The average in Eq.(\ref{eq:average}) can hence be evaluated by means of the standard relation, to be intended order by order in the perturbative expansion,
\begin{eqnarray}
\langle O\rangle_i&=&\frac{\mathrm{Tr}\left\{e^{-\beta\,(H_0+H_i)}O\right\}}{\mathrm{Tr}\left\{e^{-\beta\,(H_0+H_i)}\right\}}=\label{perturbative}\\
&=&\frac{Tr\left\{e^{-\beta\,H_0} O_{\mathrm{INT}}\left(\beta\,\right) T_\tau\left\{e^{-\int_0^{\beta} H_{i,\mathrm{INT}}(\tau)d\tau}\right\}\right\}}{Tr\left\{e^{-\beta\,{H_0}}T_\tau\left\{e^{-\int_0^\beta, H_{i,\mathrm{INT}}(\tau)d\tau}\right\}\right\}}\nonumber
\end{eqnarray}
where $T_\tau$ is the time ordering with respect to $\tau$.\\
It can be shown that $\langle\rho(x)\rangle_i=2k_F/\pi+(\Delta N_++\Delta N_-)/L$, with $i=1,2,3$, to all orders in perturbation theory, that is, the electron density is neither renormalized nor it acquires spatial oscillations due to the presence of the impurity potential. In fact all odd orders are zero due to the Klein factors, while all even orders do not conserve the number of bosonic excitations, so that they vanish under the trace. Though the result is obtained by properly considering the Klein factors, an easy, and commonly accepted, way to see why all even orders are not renormalized by the impurity is to address the bosonized form of the Hamiltonians $H_i$, $i=1,2,3$, as it emerges if Klein factors are neglected. One has\cite{helical2}
\begin{eqnarray}
H_1&=&v'_1\cos2\phi(0),\label{eq:bos1}\\
H_2&=&v'_2\cos4\phi(0),\label{eq:bos2}\\
H_3&=&v'_3\partial_x^2\theta\large|_{x=0}\cos2\phi(0),\label{eq:bos3}
\end{eqnarray}
where the new coefficients $v'_i$ can be related to the $v_i$ as specified in\cite{helical2}.
By inserting Eqs.(\ref{eq:bos1}-\ref{eq:bos3}) into Eq.(\ref{perturbative}), it is evident that, since $\phi(0)$ and $\partial_x^2\theta\large_{x=0}$ are linear in bosonic operators, and $\rho_{\mathrm{bos}}$ is also linear in the bosons, the average of the operator $\rho_{\mathrm{bos}}(x)$ is zero to all even orders. On top of that, since Eqs.(\ref{eq:bos1}-\ref{eq:bos3}) only contain bosonic operators, the impurity potentials do not renormalize the averages of the operators $N_+/L$ and $N_-/L$ appearing in the density operator. The three impurity potentials thus do not influence the local charge density, in sharp contrast to the usual Luttinger liquid case. This behavior is a genuine effect of spin momentum locking, and hence of the topological nature of the one dimensional conducting channel. Even though the spectral density can be significantly different from the particle density,\cite{secchi,njp} we argue that our prediction might be measurable in STM experiments.\\
The case of $s_z(x)$ is not different from the case of the electron density: it contains no Klein factors and, apart from the zero mode contribution $(\Delta N_+-\Delta N_-)/(2L)$, which averages to zero, is linear in the bosonic operators. Thus we find, for all impurity potentials and up to all perturbative orders, that
\begin{equation}
\langle s_z(x)\rangle_i=\frac{\Delta N_+-\Delta N_-}{2L}=0\,\,\,\,\,\,\,\,\,\, i=1,2,3.
\end{equation}
Similarly one has
\begin{equation}
\langle s_x(x)\rangle_i=\langle s_y(x)\rangle_i=0\,\,\,\,\,\,\,\,\,\, i=2,3.
\end{equation}
For the case of $i=2$ the relation can be easily shown by noticing that all perturbative orders are zero due to the Klein factors. In the case of $i=3$, on the other hand, all even orders are zero due to the Klein factors, while odd orders are zero due to the conservation of the number of bosons in $H_0$.\\
The only non-zero averages are $\langle s_x(x)\rangle_1$ and $\langle s_y(x)\rangle_1$. The problem of calculating these averages can be easily mapped onto the the problem of calculating the Friedel oscillations induced by an impurity in the particle density of a one channel (spin polarized) Luttinger liquid.
In fact the density $\rho_{\mathrm{1LL}}(x)$ of a one-channel Luttinger liquid is
\begin{eqnarray}
\rho_{\mathrm{1LL}}(x)&=&\rho_{\mathrm{LW},\mathrm{1LL}}(x)+\rho_{\mathrm{F},\mathrm{1LL}}(x),\\
\rho_{\mathrm{LW},\mathrm{1LL}}(x)&=&\psi^\dag_+(x)\psi_+(x)+\psi^\dag_-(x)\psi_-(x),\\
\rho_{\mathrm{F},\mathrm{1LL}}(x)&=&\psi^\dag_+(x)\psi_-(x)+\psi^\dag_-(x)\psi_+(x)
\end{eqnarray}
and the coupling $V_{\mathrm{1LL}}$ to a non-magnetic impurity placed at $x=0$ can hence be modeled as
\begin{equation}
V_{\mathrm{1LL}}=V_{0,\mathrm{1LL}}\rho_{\mathrm{1LL}}(0),
\end{equation}
with $V_{0,\mathrm{1LL}}$ a real parameter.
It is not difficult to show that the coupling of the impurity with $\rho_{\mathrm{LW},\mathrm{1LL}}(x)$ (which is linear in the bosonic operators) can be treated by means of a simple canonical transformation. To derive the average electron density $\langle\rho_{\mathrm{1LL}}(x)\rangle_{V_{\mathrm{1LL}}}$ in the presence of the impurity, we then have to evaluate
\begin{equation}
\langle\rho_{\mathrm{1LL}}(x)\rangle_{V_{\mathrm{1LL}}}\sim\frac {\mathrm{Tr}\left\{e^{-\beta({H_0+V_{0,\mathrm{1LL}}\rho_{F,\mathrm{1LL}}(0)})}\rho_{F,\mathrm{1LL}}(x)\right\}}
{\mathrm{Tr}\left\{e^{-\beta({H_0+V_{0,\mathrm{1LL}}\rho_{F,\mathrm{1LL}}(0)})}\right\}}.
\end{equation}
This expression formally identical to $\langle s_x(x)\rangle_1$. The term $\langle s_y(x)\rangle_1$ can be analogously derived.\\
The exact result was obtained in Ref.\cite{exactfriedel1} (see also Ref.\cite{exactfriedel2} for the refermionization solution at $K_L=1/2$): the oscillations are characterized by the wavevector $2k_F$ and decay as $x^{-K_L}$, where $x$ is the distance from the impurity. This result holds for large $x$. The result is consistent with the fact that spin oscillations can develop in magnetically defined helical quantum dots.
\section{Conclusions}
\label{sec:conclusions}
In this Letter we have inspected several local observables of the hLL. We have demonstrated that the density-density correlation function do not show oscillations, due to the absence of Friedel oscillations in the density operator. Physically this absence is due to spin-momentum locking and hence, ultimately, to the topological nature of the system. We have further derived that the the spin-spin correlation functions are strongly anisotropic and witness the formation of a complex spin structure. This behaviour has important consequences on the spin-resolved density-density correlation functions, which resemble a planar helix. Finally we have considered the effect of three possible potentials induced by a local impurity and we have shown that, unexpectedly, the electron density is not modified by the impurity. A spin wave in the $xy$ plane can instead be pinned by a magnetic impurity causing one-particle backscattering. We have argued that the spin wave is formally analogous to Friedel oscillations of the density of a regular spinless Luttinger liquid, and hence it decays as $x^{-K_L}$ for large $x$, and its wavevector is $2k_F$.

\acknowledgments
We acknowledge financial support by the DFG (SPP1666 and SFB1170 "ToCoTronics"), the Helmholtz Foundation (VITI), and the ENB Graduate school on "Topological Insulators". We thank M. Sassetti e F. Cavaliere for interesting discussions.

\end{document}